\documentclass[useAMS,usenatbib]{mn2e}
\usepackage{graphicx}
\usepackage{latexsym}
\usepackage{url}
\usepackage{psfig}

\title{The spectral energy distribution of protoplanetary diss around Massive Young Stellar Objects}

\author[Barbara Ercolano, Antonia Bevan, Thomas Robitaille]{Barbara Ercolano$^{1,2}$\thanks{E-mail: ercolano@usm.lmu.de (BE)}, Antonia Bevan$^{3}$,Thomas Robitaille$^{4}$\\
$^{1}$USM LMU\\
$^{2}$Excellence Cluster `Universe'\\
$^{3}$Institute of Astronomy, University of Cambridge\\
$^{4}$Max Planck Institute for Astronomy, K\"onigstuhl 17, Heidelberg 69117, Germany}

\begin{document}

\pagerange{\pageref{firstpage}--\pageref{lastpage}} \pubyear{2011}

\maketitle

\label{firstpage}

\def\mnras{MNRAS}
\def\nat{Nature}
\def\apj{ApJ}
\def\aap{A\&A}
\def\apjl{ApJL}
\def\apjs{ApJS}
\def\bain{BAIN}
\def\araa{ARA\&A}
\def\pasp{PASP}
\def\aj{AJ}
\def\pasj{PASJ}
\def\ga{\sim}

\begin{abstract}

We investigate the effect of ionising radiation from  Massive Young Stellar Objects impinging on their emerging spectral energy distribution. By means of detailed radiative transfer calculations including both the gaseous and dust phase of their surrounding protoplanetary discs we highlight the importance of modelling both phases simultaneously when interpreting observations from such objects. In particular we find that models that only include dust may lead to incorrect conclusions about the inner disc evolution. Furthermore the omission of gas from models overproduces far-infrared and sub-millimiter fluxes with the result that derived dust masses may be underestimated by a factor of two in some cases. Finally free-free emission from the ionised component of gaseous discs causes the slope of the dust emission in the sub-mm and mm regime to appear flatter, resulting in incorrectly modelling the dust properties, with consequences on the derived disc masses, power law index of the surface density profile and other disc properties. 

\end{abstract}


\section{Introduction}

Over the past two decades, the number of observations of protoplanetary discs has rapidly accelerated, and with it our understanding of the properties of these objects.  With advances in resolution and precision, instruments such as the Sub-Millimeter Array (SMA) and \textit{Spitzer} have allowed for the detection of many hundreds of protoplanetary discs in various environments.  In particular, large surveys performed by \textit{Spitzer} have mapped $\sim90$\% of all the star-forming regions within 500pc of the Sun and have obtained spectra for over 2000 YSOs contained therein. 

Observations of those star-forming regions nearest to us have, to date, dominated the literature as they contain predominantly low mass, comparatively isolated stars which allow the structure of discs and their evolution to be determined more easily than their higher mass counterparts.  These include regions such as Taurus, Ophiucus and Chamaeleon I. 
More recently, studies have looked slightly further afield to the Orion nebula cluster where, in addition to studies in the infrared, 
the HST has beautifully captured images of protoplanetary discs in the optical.  

To date, nearly all identified protoplanetary discs surround low or intermediate mass stars, with only a few detected around stars with masses $>8M_{\sun}$. Despite many searches, no convincing evidence for the existence of accretion discs around protostars with masses $>20$M$_{\odot}$ (O-type stars) has yet been found (e.g., Cesaroni et al. 2007). Moreover, only a very limited number of potentially disc-harbouring sources with masses $>10$M$_{\odot}$
(early B-type stars) are known. One of these is W33A with a mass of 10 to 15M$_{\odot}$ (Davies et al., 2010). A non-spherical, compact component was detected with MIDI at the VLTI (de Wit, 2010). Another example is IRAS 13481-6124 with a mass of about 20M$_{\odot}$. Kraus et
al. (2010) reported from interferometric measurements in the near-infrared, obtained with AMBER at the VLTI, a compact, elongated, hot inner component. A final example is the ’Silhouette Disk’ around a young star in M17 (Chini et al. 2004). The mass of the central object is uncertain, however, and could be as low as 3M$_{\odot}$ (Nielbock et al., 2008). No circumstellar disc surrounding a star larger than $\sim20M_{\sun}$ has so far been reported (Preibisch et al 2011).

The existence of circumstellar discs around massive young stellar objects (MYSOs) is still sometimes questioned in the literature, as it remains unclear by what process high-mass stars form.  Frequency dependent radiative feedback simulations by Kuiper et al. (2010) 
indicate that it is possible for high mass stars to form by the same mechanism as lower mass stars.  However, it is not clear how reliable these results are as the numerical resolution of the simulations is low (1.27AU).  An alternative to this process is discussed by Zinnecker and Yorke (2007) 
 whereby several low-mass stars merge together to form a high-mass star.

Even assuming the existence of discs around MYSOs, the detection of these objects is a significant challenge:  massive stars generally form at large distances, they are normally deeply embedded in the cloud in which they form and they evolve very quickly making them relatively rare. 
Additionally, the discs themselves have a very short dispersal timescale and thus MYSO-disc systems are likely very rare.

However, very recent studies by Preibisch et al. (2011) and Grellmann et al. (2011) suggest the discovery of the potential first discs around high-mass stars.  Both focus on a single disc-star system, the first in the Carina Nebula and the second just north of the Cone Nebula. Both studies fit the model SEDs from Robitaille et al. (2006) to their observations, and derive approximate stellar masses of $\sim12M_{\sun}$ and $\sim11.6M_{\sun}$ respectively.

This technique is valuable as it 
allows information about the structure, composition and evolution of discs to be discerned.  Heretofore, models used for this purpose have only needed to simulate stars at lower masses and temperatures, generally of A/F type or later, since observations were generally restricted to stars of these types.  At these temperatures ($\leq$ 
7000K), the stars have very little or no ionizing component to their energy spectrum (assuming a blackbody energy distribution).  As a result, the models for discs around low mass stars need only to consider radiative transfer via the dust particles in the discs, as the gas opacities are orders of magnitude lower than the dust opacities at non ionising frequencies. 

However, the new discovery of discs around MYSOs creates a need for more accurate models. Future observations will hopefully allow for the study of more discs around even more massive YSOs ($>20M_{\sun}$), where a significant fraction of the stellar continuum is emitted longward of the Lyman edge. For this reason, it may be necessary to consider the effect of interactions between the stellar ionizing photons and the gas component in the disc on the spectrum that the dust is exposed to, and thus on the overall SED of the disc-star system. 
With a typical disc gas-to-dust mass ratio of $\sim$100, it is likely that the ionizing component of the star's radiation field will interact with gas particles before being absorbed by any dust particles, and thus that the dust will be exposed to an overall softer radiation field than in the "dust-only" case.  This could potentially result in less emission than expected from the dust in the mid-IR region, and thus lead to an underestimate of the mass contained within the disc.  This is particularly relevant if there is any significant difference between prediction of models with and without gas at wavelengths that are frequently used for observations; we will consider differences in the models at the \textit{Spitzer} Infrared Array Camera (IRAC) wavelengths (3.6, 4.5, 5.8 and 8$\mu m$) and the \textit{Spitzer} Multiband Imaging Photometer (MIPS) wavelengths (24, 70 and 160$\mu m$), in addition to those wavelengths used to observe the circumstellar disc surrounding the Carina Nebula (see Table 1 (reference 1), Preibisch et al. 2011).  

In section 2 of this paper we describe our numerical approach and modelling strategy, in Section 3 we present our results, which are then discussed in more detail in Section 4. Finally our main results and conclusions are summarised in Section 5.

\section{Numerical approach}

We use the 3D photoionisation code MOCASSIN (Ercolano et al., 2003, 2005, 2008a) to produce the SEDs of disc-star systems for stars at temperatures of 5800K, 10000K, 20000K, 30000K, 40000K and 50000K.  The code uses a Monte Carlo approach to the transfer of radiation, allowing the treatment of both the direct stellar radiation as well as diffuse fields and the transfer through dust and gas phase. The code includes all the dominant microphysical processes that influence the gas ionisation balance and the thermal balance of dust and gas, including processes that couple the gas and dust phases. In the ionised region the dominant heating process for typical gas abundances is photoionisation of hydrogen, which is balanced by cooling by collisionally excited line emission (dominant), recombination line emission, free-bound and free-free emission. The atomic database includes opacity data from Verner et al. (1993) and Verner \& Yakovlev (1995), energy levels, collision strengths and transition probabilities from Version 7 of the CHIANTI database (Landi et al. 2006, and references therein) and the improved hydrogen and helium free-bound continuous emission data of Ercolano \& Storey (2006). The code was originally developed for the detailed spectroscopic modelling of ionised gaseous nebulae (e.g. Ercolano et al 2004, 2007), but is regularly applied to a wide range of astrophysical environments, including protoplanetary discs (e.g. Ercolano et al 2008b, 2009, Owen et al 2010, Schisano et al 2010). Arbitrary ionising spectra can be used as well as multiple ionisation sources whose ionised volumes may or may not overlap, with the overlap region being self-consistently treated by the code. Arbitrary dust abundances, compositions and size distributions can be used, with independent grain temperatures calculated for individual grain sizes, allowing to self-consistently calculate the sublimation radii of grains of different sizes. 

We use blackbody input spectra for the stellar radiation at the effective temperatures specified above and for each temperature we consider four different optical depths (where we define the optical depth at the midplane of the disc $\tau_{mp}$) varying from the optically thin model $\tau_{mp}=0.1$ to the optically thick models $\tau_{mp}=1.0,10.0$ and $100.0$. We describe the dust density distribution similarly to Pascucci et al. (2004) by:
\begin{equation}
\label{eq:disc}
\rho(r, z) = \rho_0\,\left(\frac{r}{r_0}\right)^{-\alpha}\,\exp{\left[-\frac{1}{2}\,\left(\frac{z}{h_0 \left(r/r_0\right)^{\beta}}\right)^2\right]},
\end{equation}
where $\alpha=1$, and $\beta=1.125$, leading to a surface density profile of $\Sigma(r)\propto r^{-0.125}$, the other disc parameters are described in Table 1 of Pascucci et al (2004) and include a disc inner radius of 1~AU and a disc height of 125~AU. 

We use the dust density distribution to compute the density distribution of the gas, setting the gas-to-dust mass ratio equal to 100.  We discuss the results along two different lines of sight through the disc: one nearly edge-on ($i=$77.5$^\circ$) and the other nearly face-on ($i=$12.5$^\circ$), where $i$ is the inclination of the disc.  

\begin{table*}
\begin{tabular}{ l*{11}{p{1cm}}}
\multicolumn{12}{c}{wavelength ($\mu$m)} \\
\hline
				& 1.26		& 1.62		& 2.15		& 3.6		& 4.5 		& 5.8		& 8			& 24		& 70		& 160		& 880 \\
\hline
t0.1\_T20\_iPOL	&	4.98	&	-3.59	&	2.52	&	12.8	&	7.60	&	12.5	&	18.3	&	27.8	&	4.35	&	0.35	&	0.00	\\
t0.1\_T30\_iMP	&	17.9	&	13.7	&	37.4	&	92.3	&	126	&	96.5	&	39.8	&	14.5	&	24.6	&	23.3	&	1,070	\\
t0.1\_T30\_iPOL	&	21.2	&	15.7	&	35.8	&	119	&	129	&	104	&	31.2	&	15.1	&	26.8	&	29.5	&	N/A	\\
t0.1\_T40\_iMP	&	53.5	&	59.3	&	136	&	293	&	291	&	150	&	43.4	&	8.94	&	20.3	&	33.0	&	682	\\
t0.1\_T40\_iPOL	&	54.4	&	62.7	&	148	&	268	&	291	&	171	&	42.0	&	12.3	&	21.2	&	42.6	&	N/A	\\
t0.1\_T50\_iMP	&	101	&	129	&	264	&	466	&	350	&	184	&	45.5	&	10.6	&	8.97	&	19.6	&	4,450	\\
t0.1\_T50\_iPOL	&	99.7	&	124	&	261	&	527	&	494	&	188	&	45.7	&	9.37	&	10.9	&	17.0	&	340	\\
t1.0\_T20\_iMP	&	4.45	&	-3.00	&	2.08	&	14.2	&	16.3	&	5.91	&	29.6	&	0.99	&	-1.67	&	-0.64	&	103	\\
t1.0\_T20\_iPOL	&	5.28	&	-4.39	&	0.91	&	16.1	&	8.08	&	9.05	&	27.6	&	1.57	&	-1.23	&	-3.35	&	N/A	\\
t1.0\_T30\_iMP	&	38.9	&	40.4	&	91.3	&	136	&	95.6	&	35.2	&	38.8	&	-1.10	&	-11.5	&	-6.08	&	580	\\
t1.0\_T30\_iPOL	&	36.3	&	42.3	&	93.0	&	141	&	84.7	&	29.7	&	39.4	&	-0.60	&	-11.7	&	-2.63	&	340	\\
t1.0\_T40\_iMP	&	150	&	184	&	390	&	347	&	174	&	63.6	&	33.1	&	-12.7	&	-24.3	&	-11.1	&	2,470	\\
t1.0\_T40\_iPOL	&	150	&	187	&	374	&	353	&	166	&	57.9	&	35.0	&	-12.6	&	-24.3	&	-13.9	&	1,810	\\
t1.0\_T50\_iMP	&	375	&	480	&	992	&	554	&	241	&	83.3	&	34.6	&	-20.4	&	-35.4	&	-18.1	&	1,210	\\
t1.0\_T50\_iPOL	&	353	&	497	&	1,020	&	501	&	221	&	84.0	&	34.0	&	-20.4	&	-34.2	&	-23.0	&	560	\\
t10\_T20\_iMP	&	5.86	&	-0.48	&	4.09	&	8.49	&	21.1	&	0.72	&	11.8	&	-4.28	&	-1.12	&	0.58	&	16.5	\\
t10\_T20\_iPOL	&	5.98	&	-3.02	&	3.31	&	10.1	&	22.0	&	1.76	&	13.4	&	-4.51	&	-1.36	&	-1.41	&	-32.3	\\
t10\_T30\_iMP	&	59.1	&	68.5	&	140	&	26.1	&	-3.47	&	-13.3	&	-4.48	&	-17.9	&	-13.1	&	-3.11	&	181	\\
t10\_T30\_iPOL	&	51.6	&	61.7	&	129	&	30.5	&	-1.90	&	-10.5	&	-3.31	&	-17.4	&	-12.8	&	-5.22	&	113	\\
t10\_T40\_iMP	&	215	&	267	&	508	&	581	&	-2.84	&	-17.2	&	-9.21	&	-21.0	&	-23.4	&	-10.3	&	158	\\
t10\_T40\_iPOL	&	196	&	246	&	475	&	56.9	&	-0.55	&	-16.9	&	-8.15	&	-21.2	&	-23.7	&	-10.3	&	193	\\
t10\_T50\_iMP	&	576	&	746	&	1,280	&	83.7	&	2.34	&	-30.7	&	-27.4	&	-36.5	&	-35.9	&	-17.7	&	185	\\
t10\_T50\_iPOL	&	515	&	647	&	1,200	&	87.6	&	-2.16	&	-30.3	&	-25.6	&	-36.5	&	-35.5	&	-17.7	&	116	\\
t100\_T20\_iMP	&	5.96	&	8.37	&	9.76	&	3.91	&	4.68	&	-8.43	&	-2.53	&	-4.83	&	-3.73	&	-2.14	&	-1.84	\\
t100\_T20\_iPOL	&	4.61	&	-3.51	&	6.45	&	-2.68	&	5.51	&	-4.05	&	11.2	&	-5.18	&	-3.17	&	-1.68	&	-31.4	\\
t100\_T30\_iMP	&	109	&	118	&	195	&	4.39	&	-19.5	&	-26.3	&	-17.0	&	-16.4	&	-14.1	&	-6.46	&	4.70	\\
t100\_T30\_iPOL	&	55.7	&	60.9	&	120	&	-0.43	&	-24.0	&	-24.1	&	-5.10	&	-16.5	&	-14.2	&	-5.47	&	-1.06	\\
t100\_T40\_iMP	&	501	&	531	&	877	&	11.4	&	-31.9	&	-41.4	&	-35.3	&	-33.7	&	-27.7	&	-14.5	&	29.1	\\
t100\_T40\_iPOL	&	243	&	289	&	573	&	-7.08	&	-36.0	&	-40.0	&	-26.9	&	-31.2	&	-27.6	&	-15.6	&	109	\\
t100\_T50\_iMP	&	1,280	&	1,390	&	1,980	&	23.8	&	-38.6	&	-56.5	&	-50.2	&	-45.6	&	-38.9	&	-22.3	&	56.2	\\
t100\_T50\_iPOL	&	594	&	820	&	1,380	&	3.89	&	-45.9	&	-56.0	&	-44.9	&	-43.1	&	-38.6	&	-21.9	&	79.6	\\

\end{tabular}
\caption{Difference between dust-gas and dust-only flux as a percentage of dust-only flux at specific wavelengths for each model at stellar temperatures $T_{\rm eff}\geq20000K$.  Each model is labelled as "t[optical depth]-T[stellar temperature in 1000K]-i[MP/POL]", where MP indicates the near mid-plane line of sight (77.5$^\circ$) and POL indicates the near pole line of sight (12.5$^\circ$). N/A entries signify cases with null denominator fluxes (within the Monte Carlo error). }

\end{table*}

We run two sets of models, one with only dust included in the grids and the other where both dust and gas are present. All models where run for solar metallicities. 
We consider the percentage flux differences between the ``dust-gas" cases and the ``dust-only" cases.  In particular, we consider the flux differences at the observational wavelengths used in Preibisch et al 2011 in addition to those wavelengths used by \textit{Spitzer} (IRAC \& MIPS) (see Table 1 for a summary).  Each model is defined by the optical depth, stellar temperature and line of sight and is labelled using the following convention "t[optical depth]-T[stellar temperature in 1000K]-i[MP/POL]", where MP indicates the near mid-plane line of sight (77.5$^\circ$) and POL indicates the near pole line of sight (12.5$^\circ$).  For example, the model with parameters $\tau_{mp}=10.0$, $T_{\rm eff}=20000$K and $i=12.5^\circ$ is labelled as "t10.0-T20-iPOL".  The percentage difference is given as a percentage of the "dust-only" flux. 

\section{Results}

\begin{figure*}
\includegraphics[width=0.9\textwidth]{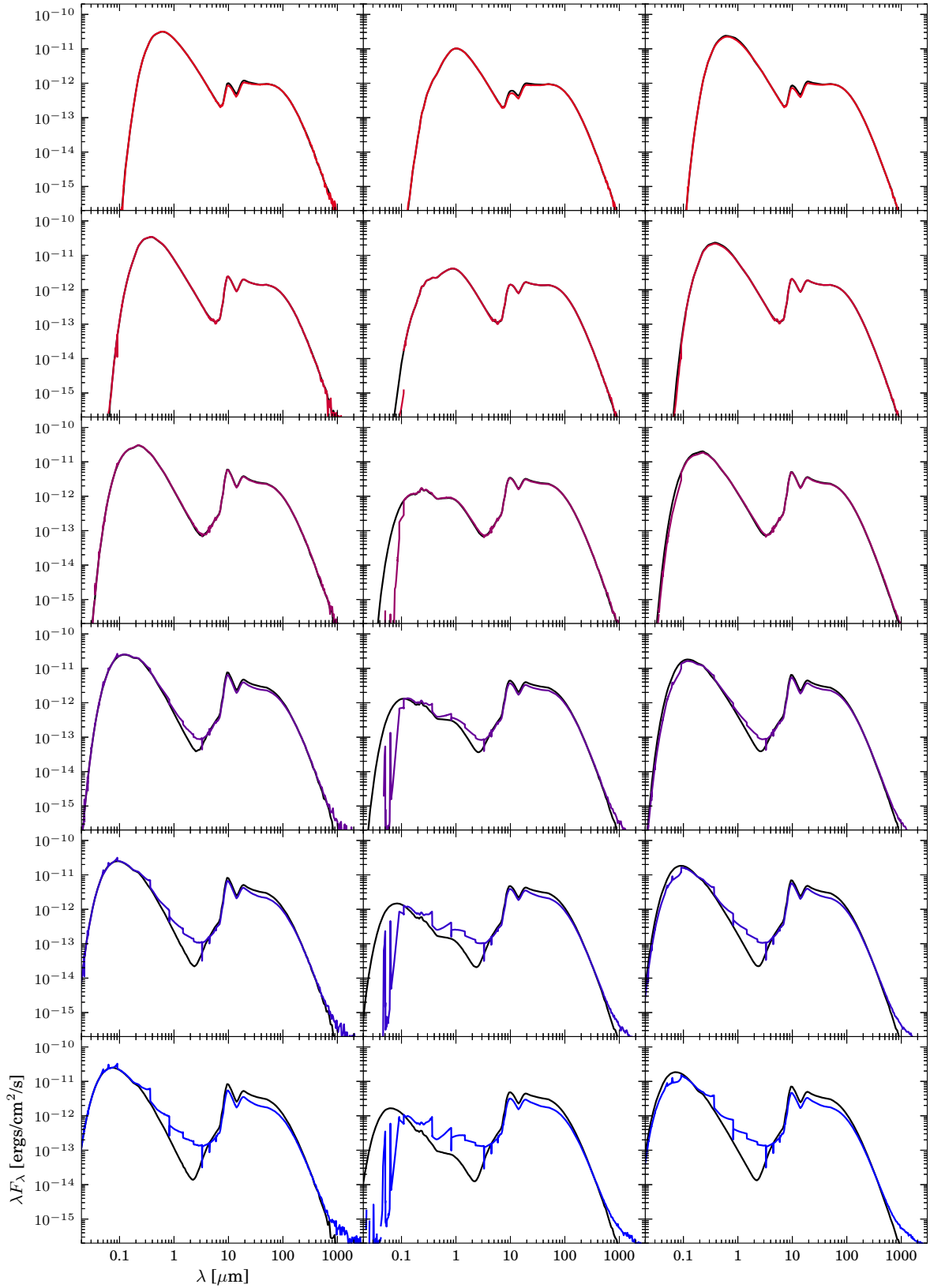}
\caption{SED plots for mid-plane optical depth $\tau_{mp}=10$.  From left to right, the SED plots are for the near-pole line of sight ($12.5^\circ$), the near mid-plane line of sight ($77.5^\circ$), and averaged over viewing angles respectively. From top to bottom, the SEDs are for stellar effective temperatures of 5800K to 50,000K. Each panel shows the SED for the dust-only case (black), and the dust-gas case (red to blue). SED plots for the $\tau_{mp}=0.1$, 1, and 100 cases are included in Appendix \ref{app:seds}.\label{fig:all_seds_10}}
\end{figure*}

\begin{figure*}
\includegraphics[width=\textwidth]{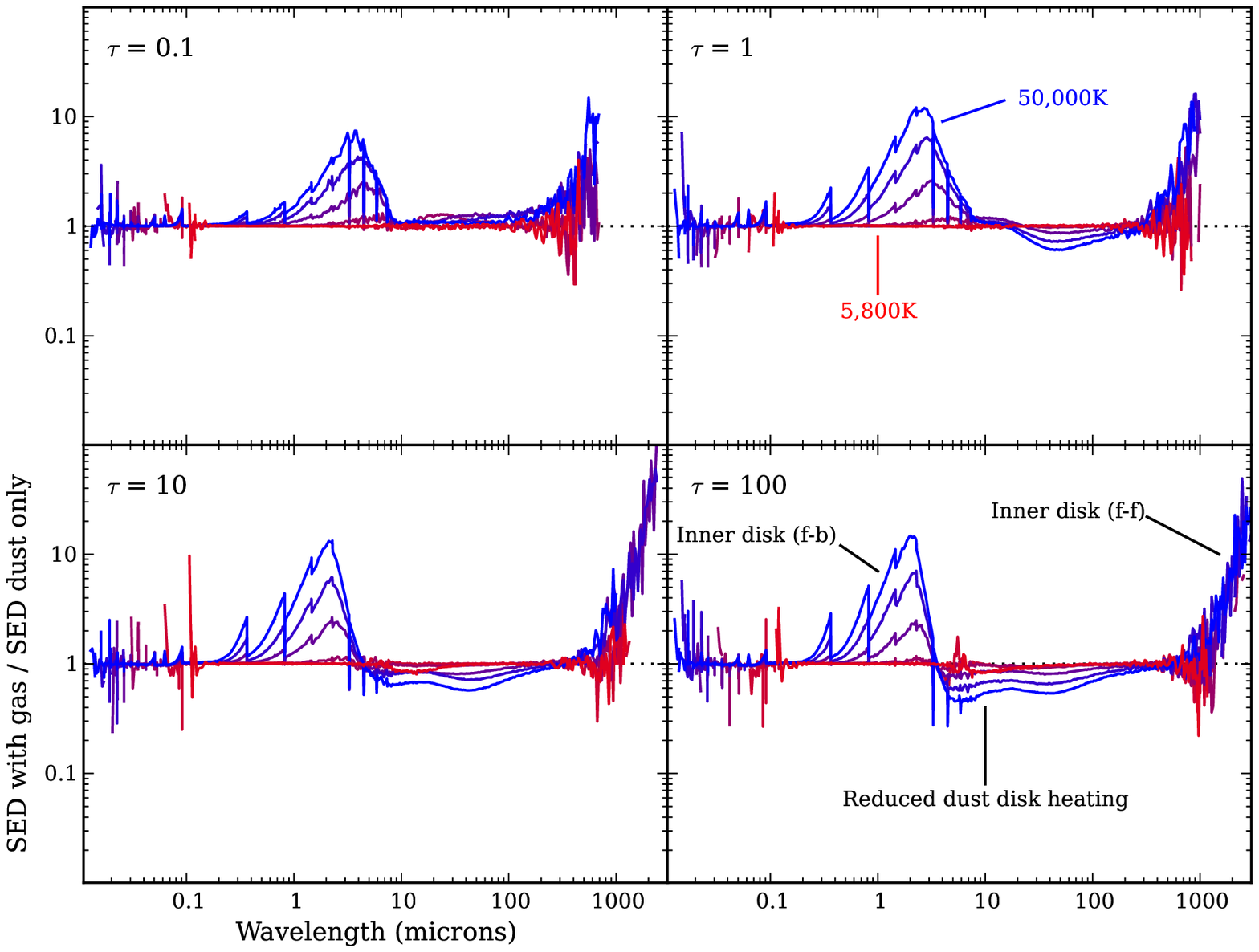}
\caption{Ratio of the dust-gas SEDs to the dust-only SEDs for the pole-on viewing angle.\label{fig:sed_ratio}}
\end{figure*}

Table 1 shows the percentual difference between the dust-gas and dust-only model fluxes at specific wavelengths for all models with T$_{\rm eff} > 20kK$. {We also include the 24 SED plots for each model in Figure \ref{fig:all_seds_10} (for $\tau_{mp}=10$, as a case study) and in Appendix A (for $\tau_{mp}=0.1$, 1, 100). We will discuss the key differences between the two sets of models according to the appearance of the SEDs at specific wavelength ranges, specifically at wavelengths $<0.1\mu m$, in the range $\sim$0.1-6$\mu m$,  in the range $\sim$6-300$\mu m$ and at wavelengths$>800\mu m$. As expected, we see no differences between the dust-gas case and the dust-only case at the two lowest effective stellar temperatures considered ($T_{\rm eff}=5800K$ and $T_{\rm eff}=10000K$), with differences first appearing at $T_{\rm eff}=20000K$, and becoming larger for larger stellar temperatures and with larger optical depth. 
This is easy to understand since only models with $T_{\rm eff} \geq 20000K$ produce enough photons at energies greater than the hydrogen ionisation potential. While these low T$_{\rm eff}$ models served their purpose of benchmarking the gas and dust transfer against the dust-only transfer, they are not useful in the context of the scientific discussion to follow and will not be further considered. 

We now consider the potential effects of ionizing photons on both the dust and the gas in the disc with reference to the differences between the dust-gas and dust-only SEDs.  We assume in our models a gas-to-dust mass ratio of 100, which results in a much higher number density of gas particles than dust particles, thus a significant percentage of the ionizing photons are absorbed by the gas particles before they have any interactions with the dust particles.  If this is the case then more ionizing stellar photons are absorbed by the disc in the dust-gas model compared to the dust-only model. This effect can be indirectly seen in Figure~\ref{fig:all_seds_10} as a reduced flux in the edge-on dust-gas case for wavelengths $<0.09\mu m$, which is due to the extinction of the light from the star along the line of sight to the observer. At all optical depths, this discrepancy can only be seen at temperatures $\geq 20000K$. Since the pole-on line-of sight does not go through the disc, the effect is not seen for that viewing angle.

In both dust-only and dust-gas cases we see thermal emission in the infrared by the dust. Since, as was mentioned above, a significant percentage of the stellar ionizing photons are absorbed by the gas particles rather than dust particles, the spectrum that the dust is exposed to is, overall, softer in the dust-gas models than in the dust-only models. Part of the stellar flux absorbed by the gas will be reprocessed to less energetic (ionising or non-ionising) continuum or lines. The total flux as a result of dust emission would therefore be reduced in the dust-gas case relative to the dust-only case, and this accounts for the differences we see in the SEDs in the wavelength range $\sim$10 - 300$\mu m$. 

Given that the gas is absorbing energetic photons, we expect to see some evidence of gas emission, which we do in the $\sim$0.1 - 10$\mu m$ wavelength range.  The shape of the curves are different in this range, with the dust-only case exhibiting a smooth curve whilst the dust-gas SEDs has the typical "step-like" appearance due to free-bound processes that dominate the gas emission coefficient at these wavelengths. As one may expect, this feature becomes more pronounced with both increasing optical depth and temperature see Figures \ref{fig:all_seds_10}, \ref{fig:all_seds_01}, \ref{fig:all_seds_1}, and \ref{fig:all_seds_100}). We note that Robitaille et al. (2006) found that the IRAC and MIPS colors of discs around massive stars for dust-only models would be very red -- the hottest dust in the disc is set by the sublimation temperature, but the stellar temperature can increase, causing a gap in wavelength space (this can be seen e.g. in Figure \ref{fig:all_seds_10}, where the bottom panels have large gaps between the photospheric and disc emission for the dust-only models). Taking into account the emission from the gas in the near- and mid-IR mitigates this effect, so that the very red colors predicted for discs by Robitaille et al. may not be seen in real MYSOs.

Finally, since a fraction of the stellar blackbody spectrum is ionising, we expect some of the gas particles to be ionised and thus for there to be free electrons present in the disc.  The flux seen at long mm wavelengths ($1000\mu m<\lambda<10000 \mu m$) in the dust-gas case is indeed due to free-free emission from the ionised gas component of the disc.  

As the effective stellar temperature increases, so does the ionizing fraction of the stellar continuum.  It is then logical to expect that all of the above effects would therefore be enhanced with increasing $T_{\rm eff}$, which is why the flux differences between the dust-gas and dust-only cases become more pronounced the higher the temperature of the central star.  

To summarize these effects, we have shown in Figure \ref{fig:sed_ratio} the ratio of the dust-gas SEDs to the dust-only SEDs.

\section{Discussion}

Multiwavelength photometry may be used to construct SEDs of observed disc-star systems.  By using fitting tools such as that detailed in Robitaille et al. (2007), a set (or sometimes multiple sets) of parameters may be identified that reproduce the observations and thus provide an insight into the physical properties of the disc-star system.  The more measurements made at distinct wavelengths the better the various physical parameters may be constrained. 

By running MOCASSIN on disc-star systems at varying stellar temperatures and optical depths considering radiative transfer through both the dust and the gas, we have shown that irradiation of the gas in discs surrounding high-mass stars has a significant effect on the overall shape of the SED.  These flux differences, particularly in certain wavelength ranges, may result in observations of massive YSO-disc systems being erroneously fitted if being fitted to "dust-only" models, such as the Robitaille et al. (2006) models. We cannot quantify this directly, since the models in this paper and the Robitaille et al. models make different assumptions in the surface density of the disc. In addition, our set of models would need to cover a wider range of parameter space in order to determine general trends. Nevertheless, we can still discuss -- qualitatively -- issues that are likely to arise if modeling discs around MYSOs with models that do not take into account the effect of the gas:



\begin{itemize}

\item Mid-infrared wavelengths probe the inner region of the disc, and therefore are very important for studying disc evolution, planet formation, etc. Especially for high-mass stars, photo-ionizing flux from the central source can ionize and evaporate the disc, starting from the inner regions (e.g. Richling \& Yorke 2000).  However, if discs around high-mass stars are modeled with radiative transfer models that do not take into account the effects of the dust, the inner disc properties could be misinterpreted, leading to incorrect conclusions about inner disc evolution.

\item The main effect of taking the gas into account on the dust in the disc is to lower the incident flux and therefore the heating, which causes the far-infrared and sub-millimetre fluxes to be reduced. Since this wavelength regime is also sensitive to the disc mass, modeling these wavelengths with dust-only models would cause the disc mass to be underestimated by more than a factor of two in some cases.

\item If one has a well-sampled SED in the sub-millimeter to mm regime, the change in slope due to the free-free emission should be unmistakable. However, in most cases, only a few fluxes are available at these wavelengths, and it is therefore possible that the change in slope would not be seen, but instead, the slope of the dust emission could appear flatter. This would result in incorrectly modeling the dust properties, which would then lead to incorrectly estimating the disc mass, disc surface density power, and other disc properties.

\end{itemize}

Though the models presented here do indeed indicate the need for high-mass YSO-disc systems to be fitted to models where gas is considered, there are a number of elements not included in our models that may further affect the shape of the SEDs.  Firstly, the presence of PAH molecules was not considered in the irradiation of these discs.  These molecules are abundant, ubiquitous and a dominant force in the interstellar medium, although little is known about their role in the structure and evolution of protoplanetary discs.  However, the presence of these molecules is inferred through strong features in the near to mid-IR.  It is possible that the inclusion of these molecules in the radiative transfer would reduce the difference in flux between the two cases in this wavelength range.

Additionally, the effects of asymmetric accretion onto the YSO may result in hot spots on the surface.  Without further modelling including a detailed accretion spectrum, it is difficult to know the effect that these hot spots would have on the shape of the SED.  In theory however, a generally cool star could have regions of its surface with significantly higher surface temperatures, emission from which could result in the ionisation of the gas in the disc (perhaps in restricted regions) and thus similar changes to the SED as those detailed above might be seen.

\section{Acknowledgements}

We thank the referee for a constructive report.

\appendix

\section{Additional SED plots}

\label{app:seds}

\begin{figure*}
\includegraphics[width=0.9\textwidth]{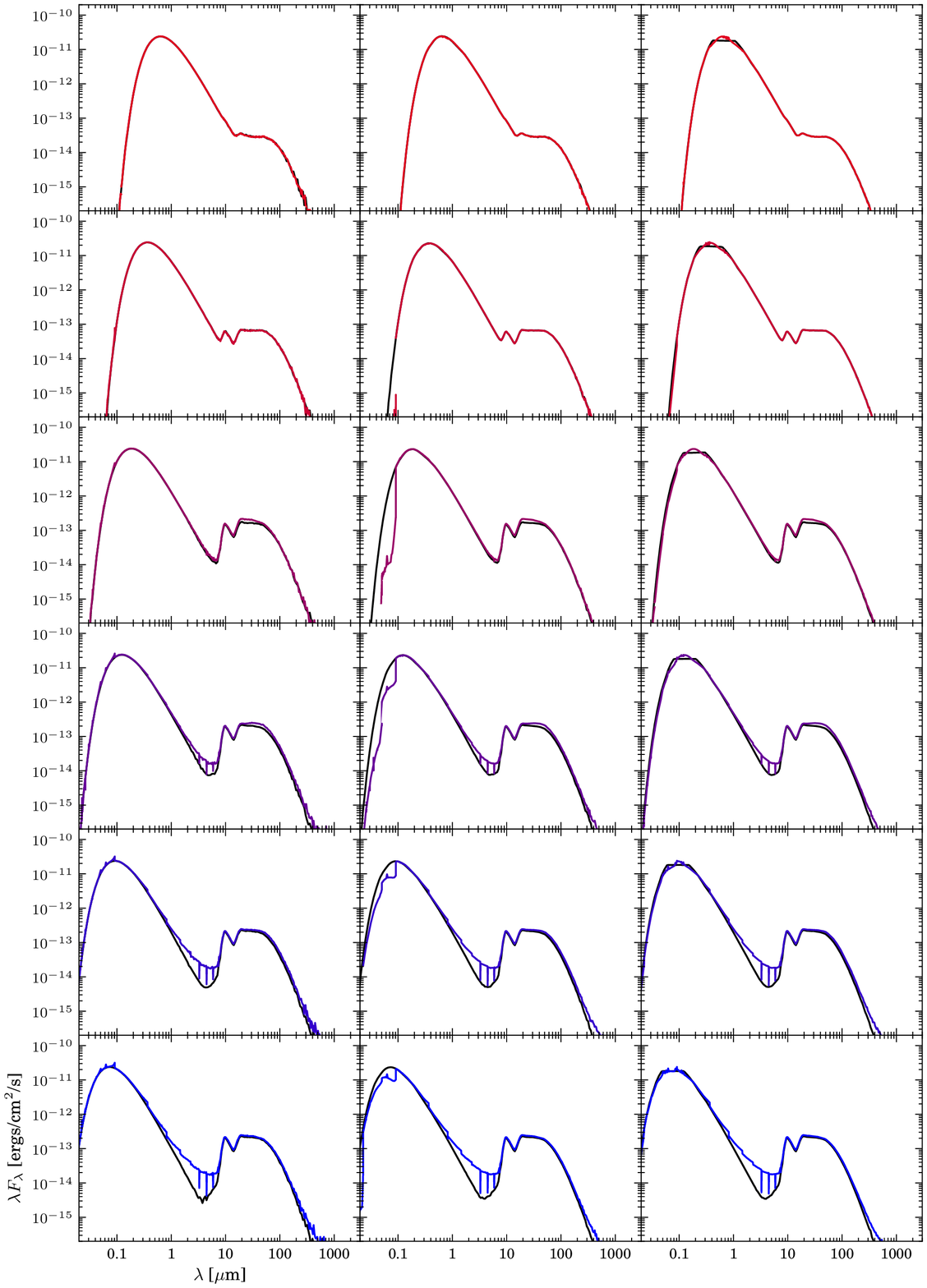}
\caption{SED plots for mid-plane optical depth $\tau_{mp}=0.1$. From left to right, the SED plots are for the near-pole line of sight ($12.5^\circ$), the near mid-plane line of sight ($77.5^\circ$), and averaged over viewing angles respectively. From top to bottom, the SEDs are for stellar effective temperatures of 5800K to 50,000K. Each panel shows the SED for the dust-only case (black), and the dust-gas case (red to blue).\label{fig:all_seds_01}}
\end{figure*}

\begin{figure*}
\includegraphics[width=0.9\textwidth]{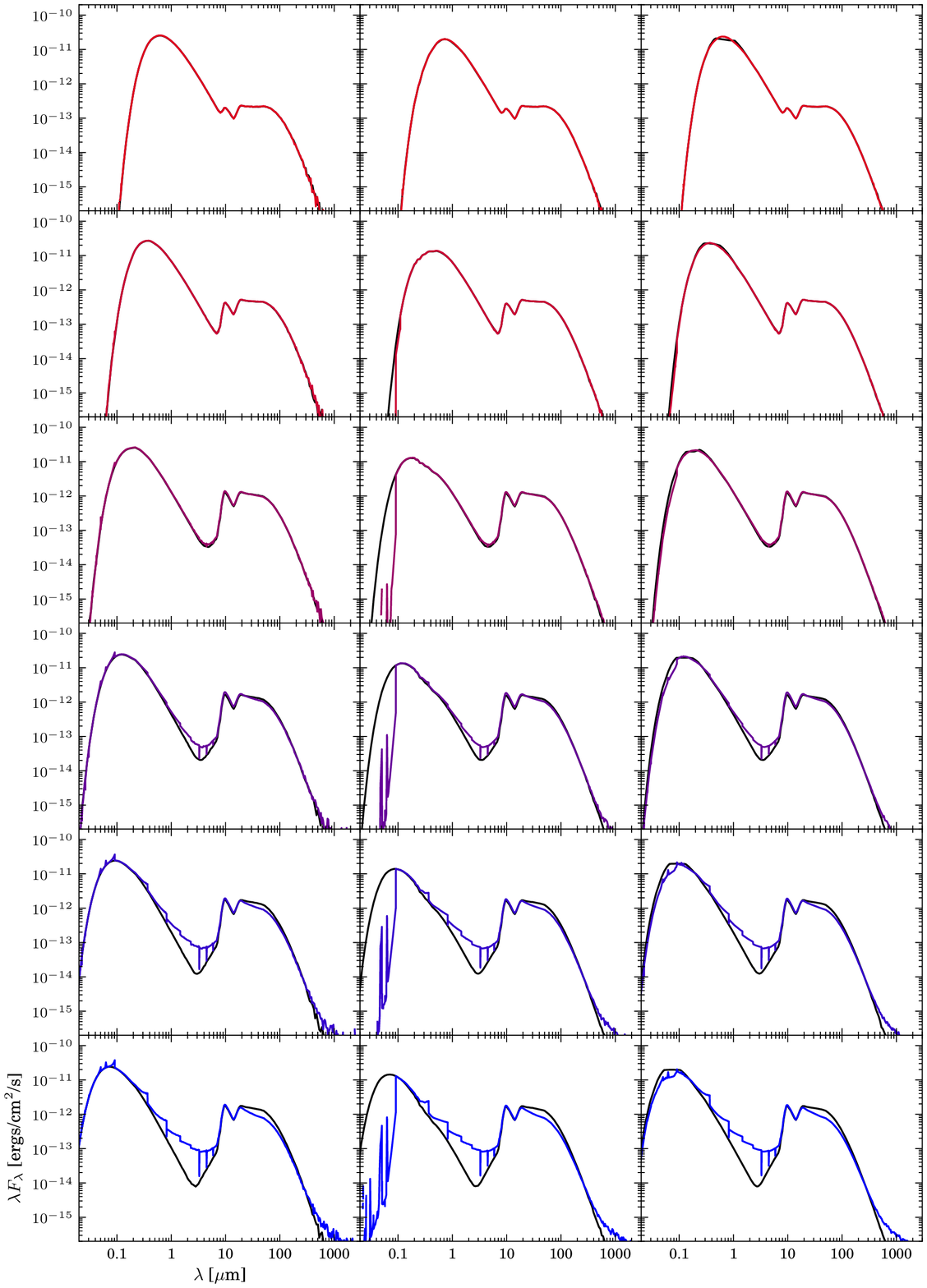}
\caption{SED plots for mid-plane optical depth $\tau_{mp}=1$. From left to right, the SED plots are for the near-pole line of sight ($12.5^\circ$), the near mid-plane line of sight ($77.5^\circ$), and averaged over viewing angles respectively. From top to bottom, the SEDs are for stellar effective temperatures of 5800K to 50,000K. Each panel shows the SED for the dust-only case (black), and the dust-gas case (red to blue).\label{fig:all_seds_1}}
\end{figure*}

\begin{figure*}
\includegraphics[width=0.9\textwidth]{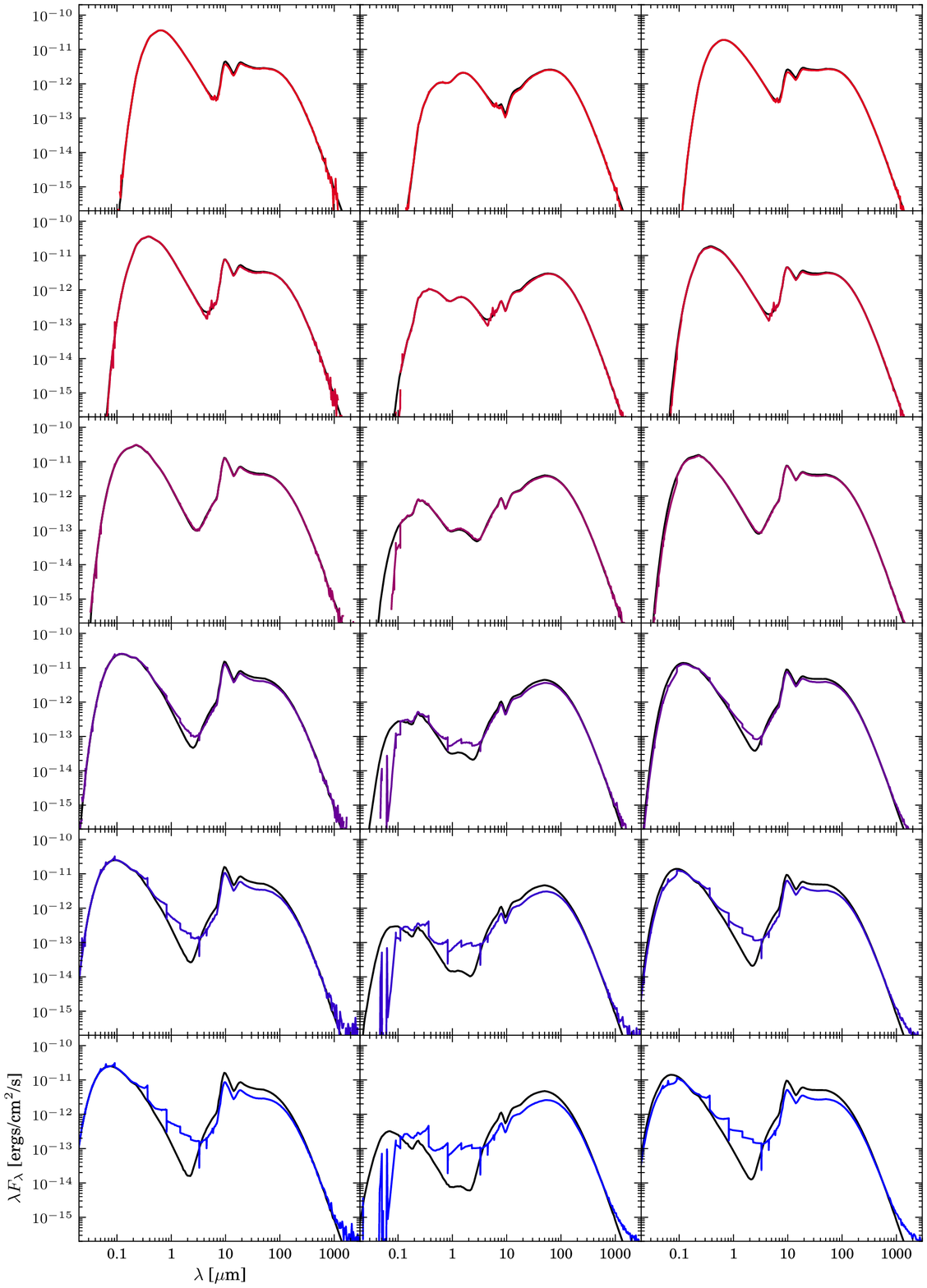}
\caption{SED plots for mid-plane optical depth $\tau_{mp}=100$. From left to right, the SED plots are for the near-pole line of sight ($12.5^\circ$), the near mid-plane line of sight ($77.5^\circ$), and averaged over viewing angles respectively. From top to bottom, the SEDs are for stellar effective temperatures of 5800K to 50,000K. Each panel shows the SED for the dust-only case (black), and the dust-gas case (red to blue).\label{fig:all_seds_100}}
\end{figure*}

\label{lastpage}

\end{document}